# Parallel, Series, and Intermediate Interconnections of Optical Nanocircuit Elements

# Part 2: Nanocircuit and Physical Interpretation

Andrea Alù, Alessandro Salandrino, and Nader Engheta[*]

University of Pennsylvania

Department of Electrical and Systems Engineering

200 South 33rd Street, Philadelphia, Pennsylvania 19104, USA

## Abstract

Applying the analytical closed-form solutions of the "quasi-static" potential distribution around two conjoined resonant half-cylinders with different permittivities, reported in the first part of our manuscript, here we interpret these results in terms of our nanocircuit paradigm applicable to nanoparticles at infrared and optical frequencies [N. Engheta, A. Salandrino, A. Alù, *Phys. Rev. Lett.* **95**, 095504 (2005)]. We investigate the possibility of connecting in series and parallel configurations plasmonic and/or dielectric nanoparticles acting as nanocircuit elements, with a goal for the design of a more complex nanocircuit circuit system with the desired response. The present analysis fully validates the heuristic predictions regarding the parallel and series combination of a pair of nanocircuit elements. Moreover, the geometries under analysis present interesting peculiar features in their wave interaction, such as the intermediate stage between the parallel and series configurations, which may be of interest for certain applications. In particular, the resonant nanocircuit configuration analyzed here may dramatically change, in a continuous way, its effective total impedance by simply rotating its orientation with respect to the polarization of the impressed optical electric field, providing a novel optical nanodevice that may alter its function by rotation with respect to the impressed optical local field.

## *1.    Introduction*

Plasmonic nanoparticles are at the basis of various anomalous optical phenomena [1]-[2]. In particular, the use of their nanoscale resonances has been recently proposed in various configurations for different purposes. Possibilities of overcoming the diffraction limitations in optical waveguiding nanostructures [2]-[6] or of inducing magnetic phenomena for designing optical negative-index metamaterials [5]-[13] are recent examples of such applications.

On a much larger scale, RF and microwave circuits, i.e., filters, transmission-lines, printed boards, etc. are at the basis of most of the electronic devices employed in our daily life. The circuit theory has been well established over the past several decades at these relatively low frequencies, and complex circuits may be easily designed and realized for different purposes. At higher frequencies, however, a straightforward scaling of circuit elements is not possible for multiple reasons, and therefore optical circuit elements in the usual sense have not been readily available in the visible regime. The extension of the circuit concepts from RF/microwave regime into such high frequencies may indeed represent unprecedented opportunities for the inherent possibility of miniaturization and speed increase in nanoelectronic devices resulting from these optical nanocircuit elements. For this purpose, recently we have introduced a paradigm for extending the circuit concepts into infrared and optical nanocircuit elements, envisioning nanocapacitors, nanoinductors and nanoresistors at infrared and optical frequencies and introducing the basic theory for their realization and utilization [11]. We have shown how this may be possible by exploiting the properties of plasmonic materials, namely noble metals, polar dielectrics [1] and some semi-conductors [14], at THz, infrared and optical frequencies, when combined with frequency windows of relatively low ohmic losses. In essence, in [11] we have shown how the design tools and mathematical machinery of circuit theory may be brought into the optical domains.

Due to the different response of metals at these high frequencies, the classic concepts of circuit elements need to be fully revisited in their use at the THz, infrared and optical frequencies. As we have shown in [11], a viable way in this direction consists of substituting the role of *conduction* current -- the one that usually flows along wires and circuit components at low frequencies -- with the *displacement* current. In this way it has been possible to introduce the concept of an equivalent impedance of nanocircuit elements, which coincides with the ratio between the voltage applied across the element volume and the displacement current flowing through it [11]. In particular, it was shown how a non-plasmonic particle (with positive real part of permittivity) effectively behaves as a nanocapacitor, a plasmonic particle (with negative real part of permittivity) may represent a nanoinductor, and the inherent losses in the materials may be seen as nanoresistor elements along the circuit.

Even though this approach allows defining voltages and currents that satisfy usual circuit Kirchhoff's voltage and current laws, important differences arise in the interaction of these nanocircuit elements with the surrounding space. In particular, the conduction current flowing in a conducting wire in low-frequency circuits is generally sustained by a motion of free charge carriers within the wire, and it is therefore confined inside those regions where such carriers are present and may freely circulate, i.e., inside the conducting materials composing the circuit. The displacement current, on the other hand, is given by the time derivative of the electric displacement $\mathbf{D} = \varepsilon \mathbf{E}$, where $\varepsilon$ is the local electric permittivity and $\mathbf{E}$ is the local electric field, and consequently this current is not confined by the circuit physical boundaries. If this property does not affect the definition of a single nanocircuit element, as done in [11], it may indeed change the way in which each nanocircuit element interacts with the others in a more complex nanocircuit architecture, and in particular affects the way in which these nanocircuit elements may be interconnected.

Despite these challenges, it is indeed possible to apply these nanocircuit concepts to plasmonic and non-plasmonic optical elements. For instance, in [3] we have envisioned optical nanotransmission lines made of nanolayers of plasmonic and non-plasmonic optical materials, showing how, following our nanocircuit paradigm, their operation is in many ways analogous to a classic transmission line at lower frequencies, composed of cascades of inductors and capacitors. In [6] we have also shown analytically how linear chains of plasmonic particles (in this analogy nanoinductors) interleaved by non-plasmonic particles or gaps (nanocapacitors) may guide energy similarly to a nanotransmission line, and in [13] we have extended similar concepts to a 3D configuration of such a nanocircuit network, envisioning a 3-D nanotransmission line optical metamaterial which may exhibit negative refraction.

Clearly, in order to extend these concepts to a more general concept of nanocircuit board, the next step consists of understanding the physics underlying the basic interconnection between a pair of nanocircuit elements, namely the possibility of having parallel or series configurations of such nanoelements. Provided that these connections are possible and that they work similarly to their low-frequency counterparts, then more complex circuits may be envisioned using such concepts. In the following, therefore, starting from the results presented in the first part of our paper for the analytical closed-form solution of the potential distribution inside and around two conjoined half-cylinders of different permittivities [16], we analyze in detail the interaction between a pair of nanocircuit elements and we show how it is possible to envision effective series and parallel configurations, in order to design the basic combinations of nanocircuit elements as the building blocks for a more complex nanocircuit. We believe that this may lead to the design of more complex circuit systems and responses at frequencies where the classic circuit concepts have not been available so far, with important applications in physics, optics, biology and other related applied fields.

Throughout the following analysis an $e^{-i\omega t}$ time dependence is assumed.

## 2. Heuristic definition of parallel and series interconnections between nanocircuit elements

One of the common properties of the conduction current ($\mathbf{J}_c = \sigma \mathbf{E}$, with $\sigma$ being the local material conductivity) and the displacement current ($\mathbf{J}_d = -i\omega \mathbf{D}$) is the continuity of their normal components at an interface between two media, guaranteed by the charge conservation and the continuity condition for $\mathbf{J}_c$ and by the continuity of the normal component of $\mathbf{D}$ at an interface for $\mathbf{J}_d$ [17]. Both these conditions are consistent with the Kirchhoff's current law in circuit theory. This similarity allows us to define the integral flux of $\mathbf{J}_d$ across one end of a nanoparticle boundary as the equivalent current circulating across the element in its nanocircuit interpretation.

As defined in [11], an isolated subwavelength nanosphere of complex permittivity $\varepsilon$, surrounded by free space with real permittivity $\varepsilon_0$ and illuminated uniformly by an impinging vertical electric field $\mathbf{E}_0$, has indeed an equivalent lumped impedance given by $Z_{sph} = \left(-i\omega\varepsilon\pi R\right)^{-1}$, defined as the ratio between the effective voltage applied at its boundaries $\langle V \rangle_{sph} = R(\varepsilon - \varepsilon_0)|\mathbf{E}_0|/(\varepsilon + 2\varepsilon_0)$ (given by the average voltage difference between the top and bottom halves of the sphere due to the scattered electric field) and the equivalent displacement current circulating across it $I_{sph} = -i\omega\varepsilon\pi R^2 (\varepsilon - \varepsilon_0)|\mathbf{E}_0|/(\varepsilon + 2\varepsilon_0)$ (defined as the integral flux of $\mathbf{J}_d$ across the particle due to the scattered electric field). As sketched in Fig. 1, the circuit is fed by an equivalent current generator, related to the impressed displacement current $I_{imp} = -i\omega(\varepsilon - \varepsilon_0)\pi R^2 |\mathbf{E}_0|$ and it closes itself on the fringe fields around the particle, that are represented by an equivalent fringe impedance $Z_{fringe} = (-i\omega 2\pi R\varepsilon_0)^{-1}$.

In particular, as reported in [11], due to the definition of $Z_{sph}$, it is evident that a dielectric nanoparticle with $\mathrm{Re}[\varepsilon] > 0$ would act as an equivalent nanocapacitor, whereas a plasmonic nanoparticle with $\mathrm{Re}[\varepsilon] < 0$ would act as a nanoinductor, both with a parallel nanoresistor that takes into account the possible presence of ohmic losses in the particles (i.e., $\mathrm{Im}[\varepsilon] > 0$). The impressed current is then divided between the impedance of the nanoelement and the fringe capacitance of the surrounding background material, as reported in Fig. 1. The possibility of combining together plasmonic and non-plasmonic particles in a judicious way, therefore, may be seen as a possibility of nanocircuit deign at optical frequencies, similar to combining together nanoinductors, nanocapacitors and nanoresistors.

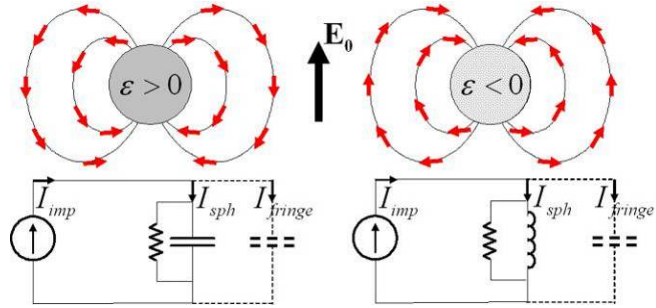

Figure 1 - (Color online). From [11]: A basic optical nanocircuit as an isolated nanoparticle illuminated by the uniform electric field $\mathbf{E_0}$, as envisioned in [11]. Left: A non-plasmonic sphere with $\mathrm{Re}[\varepsilon] > 0$, which provides a nanocapacitor and a nanoresistor. Right: A plasmonic sphere with $\mathrm{Re}[\varepsilon] < 0$, which gives a nano-inductor and a nano-resistor. The thinner field lines together with the red arrows represent the fringe dipolar electric field from the nanosphere, which corresponds to the fringe capacitor in the circuit equivalence on the bottom of each case. Copyright © 2005, American Physical Society.

To this end, we can apply the aforementioned conceptual equivalence between $\mathbf{J}_c$ and $\mathbf{J}_d$ and their agreement with Kirchhoff's current law to envision a *series* interconnection between two nanocircuit elements (we recall here that two circuit elements may be considered in series when the same amount of current flows across them). We may predict in fact that if the two nanoparticles share a common interface, with the electric displacement vector locally normal to that interface, then the displacement current $\mathbf{J}_d$ passing across one element through the common interface necessarily flows all inside the second element for the continuity condition, producing effectively a series cascade of the two nanoelements.

On the other hand, the definition of voltage applied on a given nanoparticle as the averaged line integral of the electric field across it, again consistent with the corresponding Kirchhoff's voltage law [11], allows predicting that an electric field tangential to the common interface between two nanoelements would make them effectively in *parallel* (again, we remind here that two circuit elements are considered in parallel when their terminals see the same voltage drop).

Following these heuristic considerations, Fig. 2a reports a sketch of a possible series combination of two nano-particles excited by an impinging electric field $\mathbf{E_0}$, together with their circuit equivalent, and Fig. 2b the corresponding parallel combination.

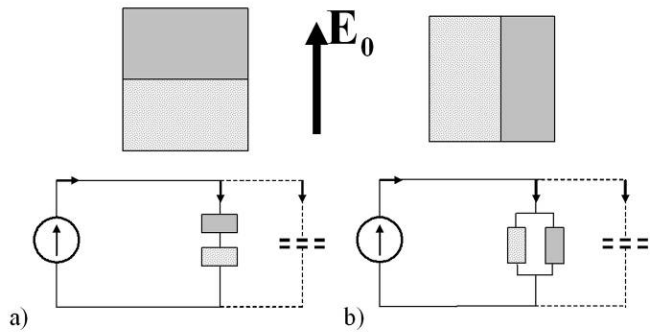

Figure 2 – A heuristic sketch of a nanocircuit series (Fig. 2a) and parallel (Fig. 2b) configurations of nanoparticles and their circuit equivalence.

It is worth noting that, unlike the classic circuit concepts, here the two lumped nanoelements may have the same structural configuration in the two combinations, and the difference in assuming their arrangement as a series or a parallel combination depends on the orientation of the exciting field. This represents an important difference with respect to the conventional circuit theory, where the lumped elements are completely isolated from the external world, and their interaction with the circuit passes only through their terminals and how they are connected with the rest of the circuit. In the configuration presented here, instead, since the fringing and external fields play a dominant role in the interaction of the nanoelement with the neighboring particles, the same connection may consist of a series or a parallel interconnection, depending on the orientation of the external applied field. This factor may be interpreted as an important new degree of freedom in optical nanocircuit design. Clearly, the configurations of Fig. 2 are two extremes of a continuous set of intermediate situations that depend on the direction of the electric field, which may be considered as a "weighted mixture" of series and parallel configurations. It must also be mentioned that we have considered scenarios in which our nanocircuit elements can be "insulated" from the surrounding space by covering the sides of each nanoelements with epsilon-near-zero (ENZ) materials as "nanoinsulators", while only the two "ends" of the nanoelement covered with epsilon-very-large (EVL) materials (as nanoconnectors) are their terminals through which the displacement current flows. These concepts have been presented elsewhere [18].

In the following sections, applying the analytical results derived in [16], we show analytically how interconnected nanoparticle configurations like those of Fig. 2 may indeed be viewed as parallel and series connections between a pair of nanocircuit elements to produce an overall response similar to the one of series and parallel lumped circuit elements in a low-frequency circuit. In this way, we want to confirm the heuristic predictions of this paragraph and open the possibility of envisioning a complex nanocircuit at optical frequencies.

### *3.     Series or Parallel L-C Configuration*

The geometry of interest, consistent with the analytical solution in the first part of our paper and with Fig. 2, is depicted in Fig. 3, in the form of two conjoined half-cylinders of radius $R$ and different permittivity are exposed to a uniform electric field, as depicted in Fig. 3. The two cases of series and parallel combinations are designed consistently with the heuristic prediction of the previous section and they are denoted for each case. Under the assumption that the size of each nanoparticle is sufficiently smaller than the operating wavelength, which is necessary to draw the circuit analogy at the core of this theory [11], we have performed our analysis in the quasi-static approximation [16].

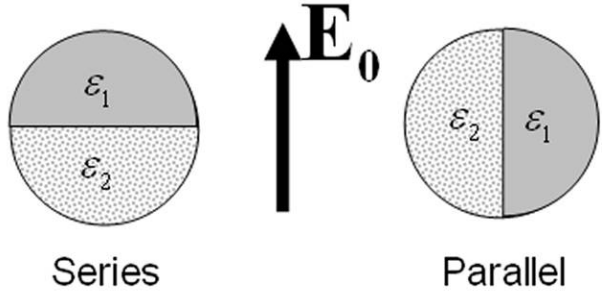

Figure 3 – Two conjoined two-dimensional half-cylinders with different permittivities, illuminated by a uniform electric field. Series (left) and parallel (right) configurations.

The distribution of the electric potential $\Phi(\rho,\varphi)$ in the two configurations of Fig. 3 may be evaluated in closed form through the solution of the Laplace equation in a suitable cylindrical reference system $(\rho,\varphi)$:

$$\rho^2 \frac{\partial^2 \Phi(\rho,\varphi)}{\partial \rho^2} + \rho \frac{\partial \Phi(\rho,\varphi)}{\partial \rho} + \frac{\partial^2 \Phi(\rho,\varphi)}{\partial \varphi^2} = 0. \tag{1}$$

This electromagnetic potential problem has been solved analytically in closed form in [16], using a Kelvin transformation to map the present geometry into a rectangular geometry and then applying a Mellin transformation of the fields in the mapped geometry. In this second part of the work, we utilize such mathematical results to show how the specific geometries of Fig. 3 may indeed be regarded as parallel or series interconnections of nanocircuit elements.

The closed-form potential distribution that we have derived for the geometries of Fig. 3 in [16] assumes a compact and relatively simple expression when the permittivities of the two half-cylinders are equal and opposite, i.e., $\varepsilon_1 = -\varepsilon_2$ and lossless scenario is assumed. As we show in the following, such a special condition corresponds to a resonant configuration. The corresponding potential expression is given by Eq. (18), (20) in [16], which we report here for convenience:

$$\Phi_i(\rho,\varphi) = \Phi_i^{parallel}(\rho,\varphi)\sin(\gamma) + \Phi_i^{series}(\rho,\varphi)\cos(\gamma), \tag{2}$$

where:

$$\begin{cases} \Phi_i^{parallel}(\rho,\varphi) = -E_0 \dfrac{\rho^2 + R^2}{\rho}\sin\varphi \\ \Phi_i^{series}(\rho,\varphi) = -E_0 \dfrac{\varepsilon_0}{\varepsilon_i} \dfrac{\rho^2 - R^2}{\rho}\cos\varphi \end{cases}, \tag{3}$$

with $R$ being the radius of the conjoined cylinder, the pedix $i = 0, 1, 2$ referring to the region where the potential is evaluated ($\Phi_i$ is evaluated in the region of permittivity $\varepsilon_i$) and $\gamma$ being the angle between the direction of polarization of the impinging electric field $\mathbf{E}_0$ and the normal to the interface between the two half-cylinders.

As evident from (2), due to the linearity of the problem the potential (2) is determined by the sum of two terms, each of them excited by the component of the impressed field parallel or orthogonal to the half-cylinder interface. Our heuristic prediction has envisioned each of these two cases corresponding to the parallel and series circuit configuration, as predicted in Fig. 2, and we verify in the following that this prediction is effectively confirmed. We note that the potential distribution in the background region $\Phi_0$ does not depend on the specific values of $\varepsilon_1$ and $\varepsilon_2$, provided that the resonant condition $\varepsilon_2 = -\varepsilon_1$, which is at the core of the validity of (3), is met. Here the two nanoelements appear to support a resonance that is independent of the relative value of their oppositely-signed permittivities. This is analogous to the case in conventional electronics where an inductance $L$ and a capacitance $C$ are connected to form a resonance at the frequency $\omega = 1/\sqrt{LC}$, while the "outer" circuit cannot distinguish the specific values of the two individual elements. Electromagnetically, the conjoined half-cylinders in this case are supporting a plasmonic resonance related to the specific geometry under analysis here and to the condition $\varepsilon_2 = -\varepsilon_1$, which requires one of the two materials to be plasmonic. We note that in this scenario the presence of a resonance at the condition $\varepsilon_2 = -\varepsilon_1$ is expected, due to the definition of impedance of the nanoelements given in the previous section. Since the two particles have the same geometry and configuration, we indeed expect the two impedances associated with the two half-nanocylinders to be the same in absolute value, but oppositely signed, when the condition $\varepsilon_2 = -\varepsilon_1$ holds. As is well known in the circuit-theory, this condition on the impedances of two circuit elements involves a resonance when they are connected with each other, consistent with our analytical findings in (2)-(3).

*a) Series resonant configuration*

In this case $\gamma = 0$ and the impinging uniform electric field is normal to the interface between the two half-cylinders (left in Fig. 3). The potential distribution in the specific case $\varepsilon_2 = -\varepsilon_1 = -2\varepsilon_0$ is reported in Fig. 4a, where the impressed electric field is vertical pointing towards the top of the figure.

As evident, the structure in this configuration perfectly confirms our heuristic prediction of a resonant L-C series configuration for the two nanocircuit elements represented by the two conjoined half-cylinders. Here losses have been neglected, and therefore we do not expect any voltage drop across the equivalent resonant L-C series pair of Fig. 2. This is reflected into the cylinder surface being perfectly equipotential as viewed from the outside, as Fig. 4a shows (we recall that in our circuit analogy the voltage drop across the pair of nanocircuit element is indeed given by the total voltage difference across the cylinder).

The equipotential lines in this configuration bend in such a way that the electric field is always locally normal to the surface of the cylinder, and thus the structure's surface has the same potential everywhere. As a result, the equivalent local impedance of the cylinder as seen from the outside is zero, implying the effective permittivity of the whole cylinder in this configuration to appear as $\varepsilon_{eff} = \infty$ as seen from the outside. Indeed no voltage drop is seen across the pair, which effectively behaves as a lumped resonant L-C serial circuit.

Analyzing the potential distribution inside the resonant cylinder the analogies with a resonant series pair is still very remarkable: inside the cylinder a voltage gradient is indeed present, very distinct from the situation of a perfectly conducting cylinder, whose potential distribution would remain constant all over the structure. Here actually the potential increases to infinity moving towards the center of the cylinder, where a double singularity is present. The impressed current entering in the lower half-cylinder, given by $-i\omega\varepsilon_0\mathbf{E}$ integrated over the lower semi-cylindrical surface is continuous at this boundary with the displacement current $-i\omega\varepsilon_2\mathbf{E}$ inside the lower half-cylinder and this is all flowing towards the upper half-cylinder through the singularity at the center. In order to have the continuity of the displacement current across the interface between the two half-cylinders, the vector $\mathbf{E}$ necessarily points downwards in the lower half-cylinder (here $\varepsilon_2 < 0$), justifying the positive (i.e., upward) gradient of the potential, and increases in magnitude moving towards the center to maintain a continuous flow of current over a narrowing region of space. The potential is therefore increasing as we approach the singular point at the origin. Here all the displacement current flows into the upper half-cylinder and from the upper cylindrical surface flows back into the outside region. All this behavior of the flow of displacement current is again consistent with the series resonant configuration in our circuit analogy, which, although looks like a short circuit as viewed from the outside circuit (i.e., no voltage drop is present at the outer terminals of the whole cylinder), there is indeed present a high amount of stored energy inside the resonant pair, and the voltage distribution varies as one "walks" inside the pair, with a maximum at the connection between the two resonant lumped elements.

Analyzing the circuit equivalence at the bottom of Fig. 2a we see that, since the series connection results in a short circuit for the outside world, all the impressed current necessarily flows inside the pair and no current passes through the fringe capacitance. This is again consistent with the results of Fig. 4a, in which all the electric field lines in the lower half space, even far from the structure, converge towards the lower half-cylinder surface and connects to the outer half space through the singular points inside the resonant cylinder. No current effectively flows to the lower to the upper half space through the space outside the cylinder, and thus there is no potential drop in the equivalent fringe capacitance in the circuit of Fig. 2a.

Swapping the two half-cylinders, i.e., having a positive-permittivity lower half-cylinder and a plasmonic upper half-cylinder for the same orientation of the impressed electric field, would not modify the outer potential distribution, but it would necessarily flip the signs of the electric field inside the two half-cylinders for the sake of continuity of the displacement current, thus reversing the sign of the potential gradient as well. This is consistent with exchanging the roles of inductance and capacitance in a resonant series L-C circuit, which flips the sign of the voltage derivative as one goes through the series cascade. Moreover, increasing (decreasing) the absolute values of permittivities, keeping the resonant condition $\varepsilon_2 = -\varepsilon_1$, would not modify the external potential distribution, as already noticed, but it would proportionally decrease (increase) the amplitude of the electric field inside the cylinder, consistent with (3) and with the continuity of the current flow. Again, this follows exactly the circuit analogy drawn in Section 2: increasing (decreasing) $|\varepsilon|$ effectively decreases (increases) the effective impedance of both resonant elements, decreasing (increasing) the voltage derivative through the pair. All these analogies between the exact closed-form solution of the scattering problem of Fig. 3 and the circuit diagram of Fig. 2a clearly confirm our heuristic predictions in this series configuration.

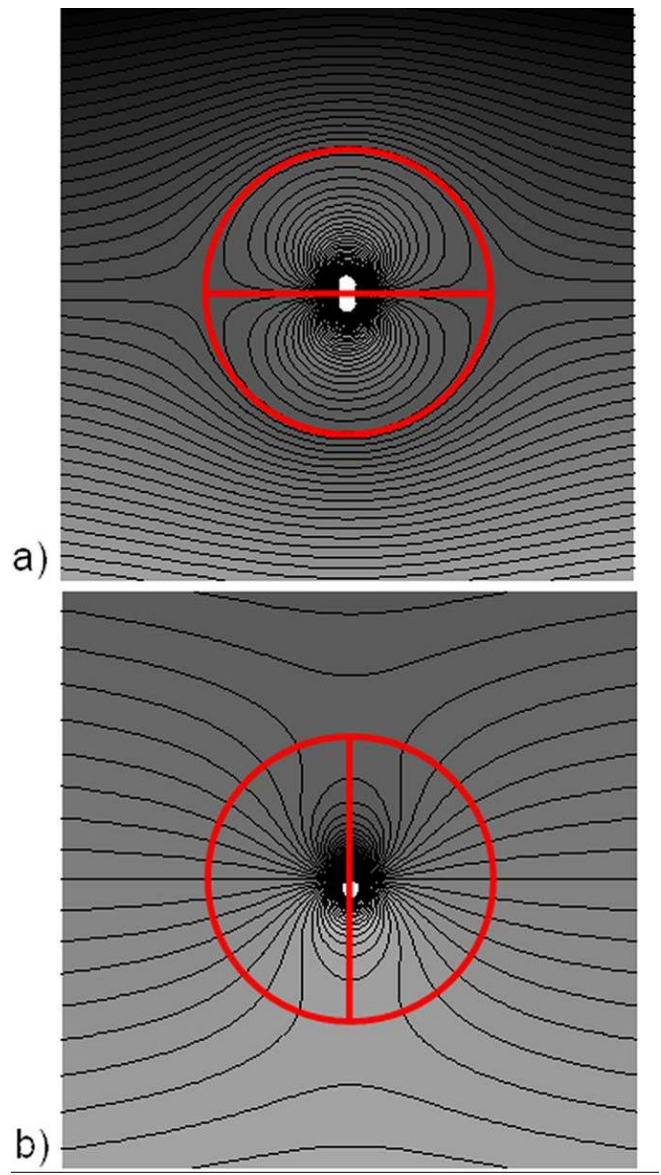

Figure 4 – (Color online). Exact "quasi-static" potential distribution for the 2D conjoined half-cylinders of Fig. 3 in the series (a) and parallel (b) resonant configurations. In both figures the uniform impressed electric field is vertical pointing to the top and for the series configuration $\varepsilon_1 = -\varepsilon_2 = 2\varepsilon_0$ (for the parallel configuration the potential distribution is independent of the value of permittivities, as long as $\varepsilon_1 = -\varepsilon_2$). Darker regions correspond to a lower potential.

*b) Parallel resonant configuration*

Consider now the case for which $\gamma = \pi/2$, i.e., the uniform applied electric field is parallel to the interface between the half-cylinders. In this case, the distribution of $\Phi^{parallel}$ is reported in Fig. 4b, with the electric field again pointing towards the top of the figure.

Inspecting Eq. (3) we notice that as long as the permittivities of the two half cylinders satisfy the resonant condition $\varepsilon_2 = -\varepsilon_1$, the potential distribution in this configuration does not depend on the specific value of $\varepsilon_1$ and $\varepsilon_2$, nor on the background material. The two conjoined half cylinders in this case behave as an isolated system, due to their intrinsic resonance.

We notice in Fig. 4b that in this case the structure indeed behaves as a parallel lossless L-C circuit at resonance, for which no net current flows in or out of the structure, as in an open circuit. In this case, the equipotential lines indeed bend in such a way that the electric field is always tangential to the surface of the cylinder, and therefore the equivalent impedance of the cylinder becomes locally infinite, implying an effective homogeneous permittivity $\varepsilon_{eff} = 0$, as seen from the outside. Moreover, no displacement current $\mathbf{J}_d$ flows in or out of the surface of the cylinder, clearly resembling the lumped parallel resonant circuit of Fig. 2b. All the impressed displacement current outside of the resonant cylinder flows into the upper half space, i.e., through the fringe capacitance in our circuit analogy of Fig. 2b.

We notice that there is indeed a displacement current flowing within the cylinder, just between the two resonant halves, through the dipolar singularity at the center of the structure. This is again analogous to what happens in a resonant lossless L-C parallel configuration, with a relevant stored electric and magnetic energy in the resonant pair and a strong circulation of current exchanged between the two resonant elements, while no net current flows in or out of the pair. We know how in such a resonant circuit pair varying the values of the two impedances or flipping their signs (i.e., changing the respective position of L and C) would induce a variation in the sign and amplitude of the circulating current. This is not directly seen in the potential distribution of Fig. 4b, which as mentioned above, remains the same, independent of the specific value of $\varepsilon_2 = -\varepsilon_1$. This is due to the fact that in this case a change in the value of $\varepsilon_i$ would indeed affect the displacement current amplitude $-i\omega\varepsilon_i \mathbf{E}$, but the potential distribution would remain unaffected, both in the geometry of Fig. 4b and in the corresponding circuit of Fig. 2b. All these considerations clearly validate our heuristic predictions on the parallel connection of two nanocircuit elements presented in Section 2.

*c) Electromagnetic properties of the resonant conjoined half-cylinders and their potential applications*

The properties depicted in this section are quite striking, if we consider that they are exact closed-form solutions of the scattering problem solved in [16]. Here we show other peculiar electromagnetic features of these resonant configurations and their possible applications.

Since these resonant configurations are independent of the specific values of the permittivities used for half cylinders as observed from the outside, as a special case we may obtain the same phenomena for a half-cylinder with permittivity $-\varepsilon_0$. Such a structure would have very interesting scattering properties, for which the direction of the impressed electric field would determine its effective permittivity in its interaction with the outside world. This is because effectively such a half-cylinder may interact with the free-space half-cylinder "connected" to it. The exact potential distribution in this specific case is shown in Fig. 5 for the series and parallel configuration (again, the impressed electric field is vertically polarized in both figures). As an aside, it should be mentioned that the parallel configuration shows the exact same potential distribution of Fig. 4b, consistent with the previous discussion, although here effectively only a half-cylinder is present.

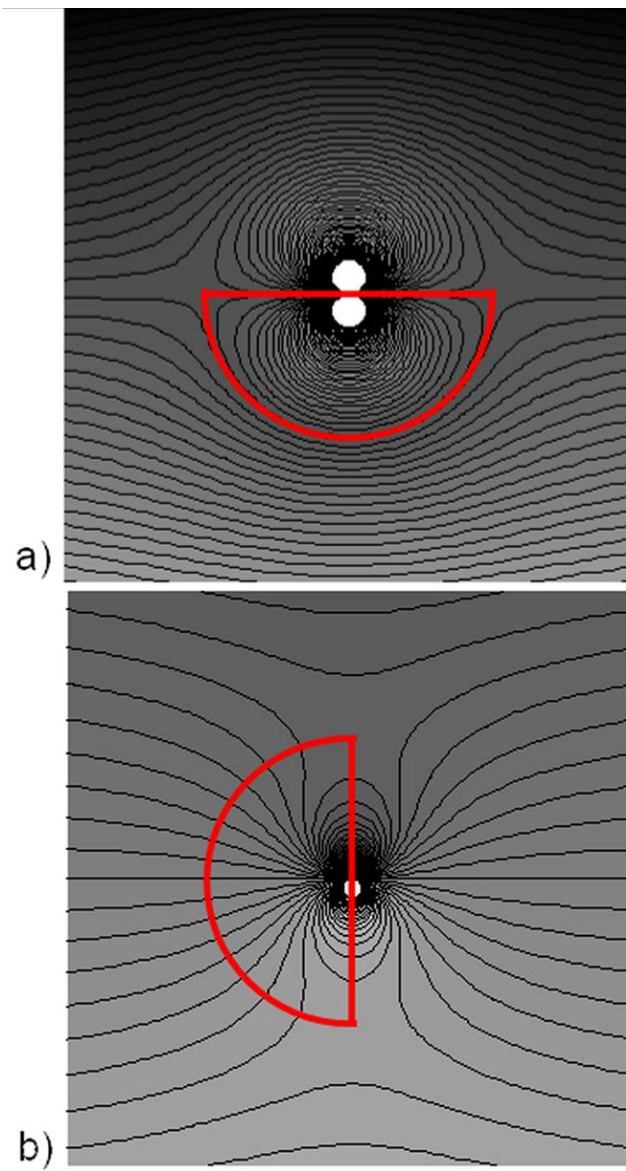

Figure 5 – (Color online). Exact "quasi-static" potential distribution for a 2D half-cylinder with $\varepsilon = -\varepsilon_0$, consistent with the results of Fig. 4. In this case, the series or parallel combination is interestingly still present, even though the other nanocircuit element that supports the resonance is represented by the free-space half-cylinder "connected" to it.

This anomalous effect may be seen as another consequence of the anomalous features of wave interaction with a plasmonic material "matched" with free space, of which the "perfect" lens is another example [19].

Following (2), for the general configuration of the two resonant half-cylinders, there is a continuous set of potential distributions evolving from the series ($\gamma = 0$) to the parallel ($\gamma = \pi/2$) configuration, in which our pair of nanocircuit elements may be interconnected. This continuous set is spanned through the variation of orientation of the impinging electric field. As an example, in Fig. 6 we illustrate the case of a "half-series-half-parallel" configuration obtained when $\gamma = \pi/4$. Again in the figure the impressed electric field is directed towards the top of the figure and one may appreciate how the equipotential lines approaching the cylinder surface now form a specific angle with the normal to the surface. Taking the gradient of (2) we find:

$$\left.\frac{E_\varphi}{E_\rho}\right|_{\rho=R} = \left.\frac{\partial \Phi_i^{parallel}(\rho,\varphi)/\partial \varphi}{\partial \Phi_i^{series}(\rho,\varphi)/\partial \rho}\right|_{\rho=R} = \tan\gamma, \qquad (4)$$

which ensures that the angle that the local electric field forms with the normal to the cylinder surface is constant and equal to $\gamma$, which is the angle between the impressed electric field and the normal to the interface between the two half-cylinders in this resonant configuration. This exact result confirms analytically the previous predictions regarding the electromagnetic behavior of the resonant cylinder in the series and parallel configurations (with electric field respectively normal and tangential to the cylindrical surface point by point) and it provides a new

"intermediate stage" of boundary condition for this resonant configuration, represented by (4). In other words, for a uniform vertical electric field a rotation of the resonant conjoined half-cylinders provides a smooth transition, following (4), from an effective perfect electric cylinder ($\gamma = 0$) to an effective perfect magnetic cylinder ($\gamma = \pi/2$), coinciding respectively with the two extremes of a short (series) and an open (parallel) circuit in our circuit analogy. This "intermediate configuration" is not available in the conventional circuits in the RF and microwave regimes, but it can be present in the plasmonic nanocircuits we envision here.

In the general case the interconnection between the two half-cylinders is neither a parallel nor a series, even though the conjoined elements are still at resonance. In the case of Fig. 6, for instance, half of the current flows through the resonant pair and the other half flows through the fringe capacitance, as if the nanocircuit pair were "matched" with free space. Since this behavior is independent of the background properties, but it only depends on the orientation of the field with respect to the interface between the two conjoined elements, this special resonant configuration represents an interesting novel "nanocircuit element" (or nanocircuit configuration) matched with any external circuit to which it is connected! The possibilities of weighting the response of the nanopair by varying the angle $\gamma$ may provide interesting new degrees of freedom in nanocircuit design, as this example shows, with various potential applications. For example, this resonant configuration may act as an optical switch, varying drastically its effective permittivity from 0 (parallel) to $\infty$ (series) by just changing its orientation with respect to the external field. In other words, this optical nanoswitch determines the fraction of current flowing through it and through the surrounding space independent of the background permittivity, i.e., independent of what "loads" the external circuit, but only as a function of its orientation with respect to the impressed field. The extreme cases of open circuit (zero current flowing through the pair) and short circuit (all the current flowing through the pair) are achieved by simply rotating the switch by 90° and the fraction of power flowing through the element is varied linearly with $\tan\gamma$, as (4) ensures. This may have other exciting potentials that are beyond the scope of the present paper.

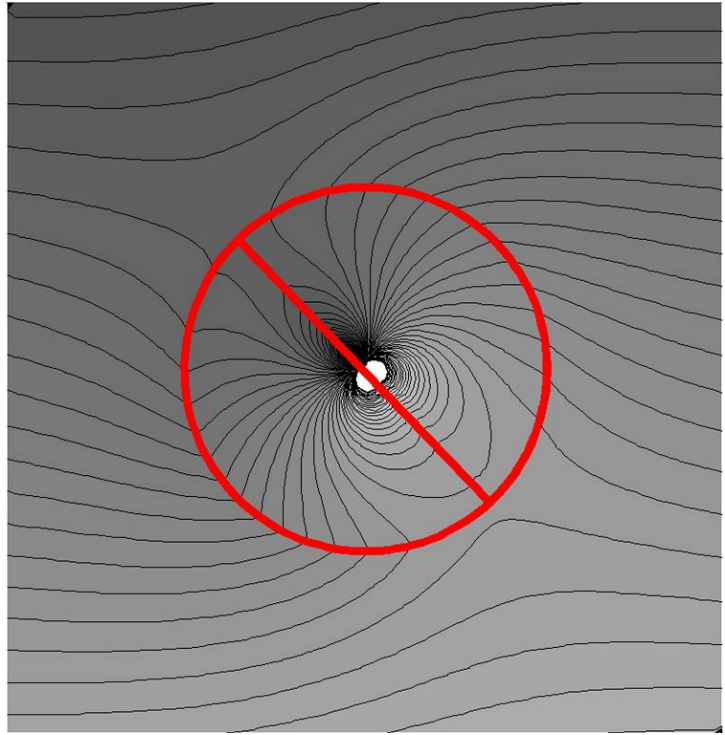

Figure 6 – (Color online). "Half-series-half-parallel" intermediate interconnection obtained from the structure of Fig. 6 when $\gamma = \pi/4$ (the uniform impressed electric field is vertical pointing to the top).

The analysis developed in this section, apart from confirming the heuristic predictions of how nanocircuit elements may be interconnected in series and parallel configuration, has also discovered interesting resonant properties of pairs of properly connected nanoparticles, which may have interesting applications in several fields. We note the analogies between these considerations on the scattering properties of this resonant pair of nanocircuit elements and the analogous behavior of the rectangular double wedge we have analyzed in [16]. Clearly, these two geometries are strictly related to each other through the Kelvin transformation. Both geometries show anomalous features that will be analyzed in more details in a future contribution.

### *4.    Non-Resonant Configurations*

In Fig. 7 we report some numerical examples for non-resonant series and parallel interconnection of conjoined half-cylinders, again applying the theory developed in [16]. In this specific case $\varepsilon_1 = 2\varepsilon_0$ and $\varepsilon_2 = -3\varepsilon_0$.

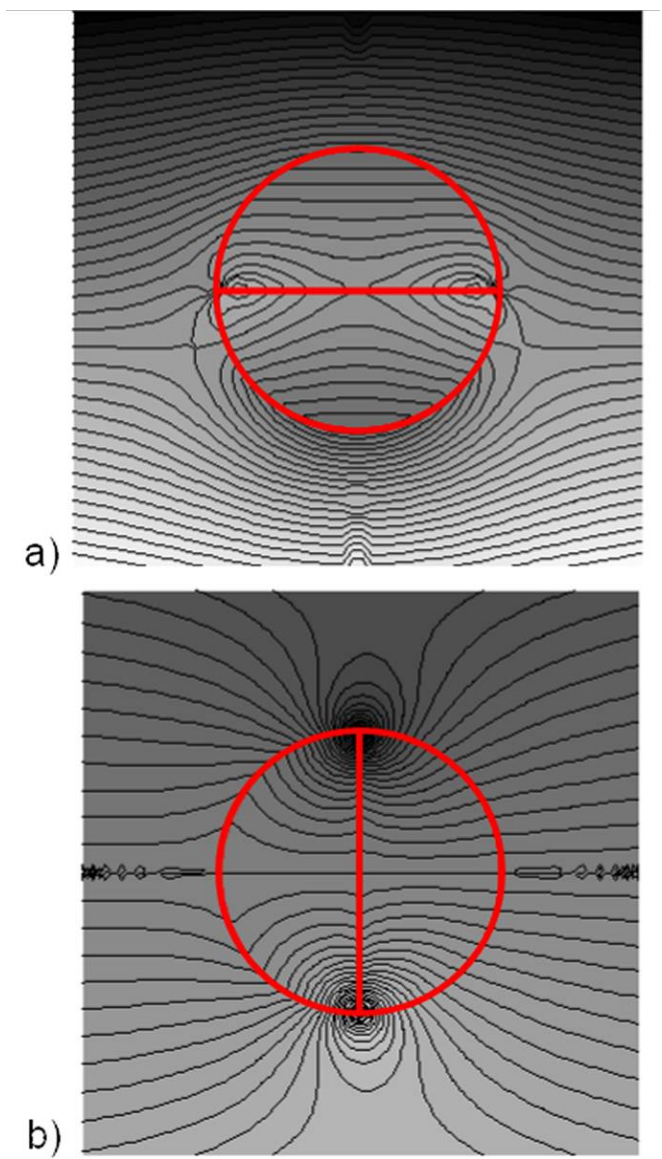

Figure 7 – (Color online). "Quasi-static" potential distribution for two 2D conjoined half-cylinders with $\varepsilon_1 = 2\varepsilon_0$, $\varepsilon_2 = -3\varepsilon_0$ in their series (a) and parallel (b) interconnection. The small numerical noise along the horizontal axis in (b) is due to the small truncation errors.

Even though the pair of nanoelements is non resonant in this case, the diagrams confirm the heuristic expectations in terms of our nanocircuit analogy. In Fig. 7a the series configuration ensures that the current entering in the lower half-cylinder flows all inside the upper half-cylinder. Due to the presence of a nanoinductance and a nanocapacitance, even though not at resonance, there is a maximum voltage at the interface between the two half-cylinders, consistent with the behavior of a series non-resonant L-C circuit. In this case, however, part of the impressed current flows to the upper half-space through the exterior region, the electric field is not all orthogonal to the cylinder surface and the equipotential lines do not close all at the cylinder surface. This is due to the non-resonant behavior, which allows some of the current to flow in the fringe capacitance of Fig. 2a.

In Fig. 7b similar considerations may be made for the parallel configuration, where the equipotential lines are orthogonal to the interface between the two half-cylinders, ensuring the same voltage drop at the terminals of the two nanocircuit elements. In this case some current can flow inside the non-resonant pair, as expected from the circuit analogy, and as verified by the equipotential lines of Fig. 7b.

As a final remark, we notice that in all these examples we verify the presence of a saddle point for the potential in the region where the three dielectric materials share a connecting point. The boundary condition on the continuity of the displacement current at that point, as it has been detailed in [16], implies indeed the presence of this saddle point in all the considered configurations.

## *5.  Conclusions*

In this contribution we have analyzed in details the optical response of a pair of two-dimensional electrically small conjoined half-nanocylinders of different permittivity under a quasi-static excitation. Relying on the analytical closed-form solution of the problem that we have developed in the first part of this paper [16], here we have analyzed in detail the case of the plasmonic nanoresonances of this geometry. They fully confirm our predictions regarding the possibility of envisioning such composed systems in terms of our nanocircuit concept, and open the path for connecting in parallel and series configurations nanocircuit elements at infrared and optical frequencies, a first step towards the design of more complex optical nanocircuit systems. Our analysis has also provided the design of a novel optical resonant nanoelement that changes, in a continuous fashion, its effective impedance with its orientation with respect to the external field polarization.

*Acknowledgements*

This work is supported in part by the U.S. Air Force Office of Scientific Research (AFOSR) grant number FA9550-05-1-0442. Andrea Alù was partially supported by the 2004 SUMMA Graduate Fellowship in Advanced Electromagnetics.

[10] G. Dolling, C. Enrich, M. Wegener, C. M. Soukoulis, and S. Linden, "Simultaneous Negative Phase and Group Velocity of Light in a Metamaterial," *Science* **312**, 892-894 (2006).

[11] S. Zhang, W. Fan, K. J. Malloy, S. R. J. Brueck, N. C. Panoiu, and R. M. Osgood, "Demonstration of metal–dielectric negative-index metamaterials with improved performance at optical frequencies," *J. Opt. Soc. Am. B* **23**, 434-438 (2006).

[12] A. N. Grigorenko, A. K. Geim, H. F. Gleeson, Y. Zhang, A. A. Firsov, I. Y. Khrushchev, and J. Petrovic, "Nanofabricated media with negative permeability at visible frequencies," *Nature* **438**, 335-338 (2005).

[13] A. Alù and N. Engheta, "Three-Dimensional Nanotransmission Lines at Optical Frequencies: a Recipe for Broadband Negative-Refraction Optical Metamaterials," *Phys. Rev. B* **75**, 024304 (2007).

[14] N. Engheta, A. Salandrino, and A. Alù, "Circuit Elements at Optical Frequencies: Nano-Inductors, Nano-Capacitors and Nano-Resistors," *Phys. Rev. Lett.* **95**, 095504 (2005).

[15] J. Gómez Rivas, C. Janke, P. Bolivar, and H. Kurz, "Transmission of THz radiation through InSb gratings of subwavelength apertures," *Opt. Express* **13**, 847-589 (2005).

[16] A. Salandrino, A. Alù, and N. Engheta, "Parallel, Series, and Intermediate Interconnections of Optical Nanocircuit Elements - Part 1: Analytical Solution," submitted for publication to this journal, together with this present manuscript.

[17] J. D. Jackson, *Classical Electrodynamics* (Wiley, New York, USA, 1999).
-26-